\RequirePackage{fix-cm}

\documentclass[smallextended,epjc3]{svjour3}       \smartqed  \usepackage{graphicx}
\usepackage{ifthen}
\usepackage{amssymb}
\usepackage{graphicx}
\usepackage{xspace}
\usepackage{hyperref}
\usepackage{fullpage}
\usepackage{cancel}
\usepackage{enumerate}
\usepackage{color}
\usepackage{mciteplus}
\usepackage[all]{hypcap} \usepackage{lineno}
\begin{document}

\def\ImprovementPDFOnly{$1.3$}
\newboolean{articletitles}
\setboolean{articletitles}{true} \newcommand{\Zg}{\mbox{$Z/\gamma^*$}\xspace}
\newcommand{\pT}{\mbox{$p_{T}$}\xspace}
\newcommand{\pTl}{\mbox{$p_{T}^{\ell}$}\xspace}
\newcommand{\pTnu}{\mbox{$p_{T}^{\nu}$}\xspace}
\newcommand{\pTW}{\mbox{$p_{T}^W$}\xspace}
\newcommand{\pTZ}{\mbox{$p_{T}^Z$}\xspace}
\newcommand{\mW}{\mbox{$m_{W}$}\xspace}
\newcommand{\mT}{\mbox{$m_{T}^{W}$}\xspace}
\newcommand{\yW}{\mbox{$y_{W}$}\xspace}
\newcommand{\deltaPDF}{\mbox{$\delta_{\rm PDF}$}\xspace}
\newcommand{\deltaCalib}{\mbox{$\delta_{\rm calib}$}\xspace}
\newcommand{\deltaStat}{\mbox{$\delta_{\rm stat}$}\xspace}
\newcommand{\deltaExp}{\mbox{$\delta_{\rm exp}$}\xspace}
\newcommand{\MET}{\mbox{$\cancel{\it{E}}_{T}$}\xspace}
\newcommand\T{\rule{0pt}{2.6ex}}       \newcommand\B{\rule[-1.2ex]{0pt}{0pt}} \newcommand\LabelGP{\mbox{$\mathbf{G^+}$}\xspace}
\newcommand\LabelGM{\mbox{$\mathbf{G^-}$}\xspace}
\newcommand\LabelLP{\mbox{$\mathbf{L^+}$}\xspace}
\newcommand\LabelLM{\mbox{$\mathbf{L^-}$}\xspace}
\newcommand\DeltaSets{\mbox{$\Delta_{\rm sets}$}\xspace}
\newcommand\ZtoMuMu{\mbox{$Z/\gamma^* \rightarrow  \mu^+\mu^-$}\xspace}

\title{Prospects for improving the LHC W boson mass measurement with forward muons}

\author{
Giuseppe Bozzi\thanksref{e1,addr1} 
\and Luca Citelli\thanksref{e2,addr2} 
\and Mika Vesterinen\thanksref{e3,addr3} 
\and Alessandro Vicini\thanksref{e4,addr4} 
}

\thankstext{e1}{email: giuseppe.bozzi@mi.infn.it}
\thankstext{e2}{email: luca.citelli@studenti.unimi.it}
\thankstext{e3}{email: mika.vesterinen@cern.ch}
\thankstext{e4}{email: alessandro.vicini@mi.infn.it}

\institute{
Dipartimento di Fisica, Universit\`a degli Studi di Milano and INFN, Sezione di Milano, Via Celoria 16, 20133 Milano (Italy) \label{addr1}
\and Dipartimento di Fisica, Universita degli Studi di Milano, Via Celoria 16, 20133 Milano (Italy) \label{addr2}
\and Dipartimento di Fisica, Universita degli Studi di Milano and INFN, Sezione di Milano and TIF Lab, Via Celoria 16, 20133 Milano (Italy) \label{addr4}
\and 
Physikalisches Institut, Ruprecht-Karls-Universitat, Im Neuenheimer Feld 226 69120 Heidelberg (Deutschland) \label{addr3}
}

\maketitle

\newcommand{\avnote}[1]{\note{av}{#1}{red}}

\begin{abstract}
Measurements of the $W$ boson mass are planned by the ATLAS and CMS experiments,
but for the time being, these may be unable to compete with the current world average
precision of 15~MeV, due to uncertainties in the PDFs.
We discuss the potential of a measurement by the LHCb experiment based on the charged lepton transverse momentum 
$p_T^{\ell}$ spectrum
in $W \to \mu\nu$ decays.
The unique forward acceptance of LHCb means that the PDF uncertainties would be
anti-correlated
with those of $p_T^{\ell}$ based measurements by ATLAS and CMS.
We compute an average of ATLAS, CMS and LHCb measurements of $m_W$ from the $p_T^{\ell}$ distribution.
Considering PDF uncertainties, 
this average is a factor of 1.3 more precise than an average of ATLAS and CMS alone.
Despite the relatively low rate of $W$ production in LHCb, we estimate that with the Run-II dataset,
a measurement could be performed with sufficient experimental precision to exploit this 
anti-correlation in PDF uncertainties.
The modelling of the lepton-pair transverse momentum distribution in the neutral current Drell-Yan process 
could be a limiting factor of this measurement and will deserve further studies.
\end{abstract}

\section{Introduction}

The Standard Model precisely relates the mass of the $W$ boson to the more precisely 
measured $Z$ boson mass, fine structure constant and Fermi constant.
The resulting indirect constraint on the $W$ boson mass from a global fit~\cite{Baak:2014ora,Awramik:2003rn,Degrassi:2014sxa}
to experimental data is roughly a factor of two more precise than the direct measurement, $\mW = 80.385 \pm 0.015$~GeV~\cite{PDG2014}, leaving room for new physics, for example in supersymmetry~\cite{Heinemeyer:2006px}.
The world average for \mW is dominated by measurements from the Fermilab Tevatron collider
experiments, CDF~\cite{Aaltonen:2013vwa,Aaltonen:2012bp} and D0~\cite{D0:2013jba,Abazov:2012bv}.
The CDF and D0 measurements used 2.1~fb$^{-1}$ and 4.9~fb$^{-1}$, respectively, out of the roughly $10$~fb$^{-1}$ Tevatron Run-II dataset.
Updates from both experiments are therefore highly anticipated.
The current measurements are mostly limited by statistical uncertainties; 
either directly through limited $W$ samples or indirectly through limited calibration samples.
The uncertainty due to the parton distribution functions (PDFs) is around 10 MeV,
with some variation between experiment, lepton flavour and fit variable.
The Tevatron measurements used three different fit variables to extract \mW:
\begin{enumerate}
\item The transverse mass, $\mT = \sqrt{2\pTl \MET(1-\cos\phi)}$,
where \pTl is the charged lepton transverse momentum, \MET is the missing transverse energy
measured by the calorimeter, which estimates the neutrino transverse momentum; 
and $\phi$ is the azimuthal opening angle between the neutrino and charged lepton.
\item The charged lepton transverse momentum \pTl itself,
\item The missing transverse energy \MET itself.
\end{enumerate}
The statistically most sensitive variable is \mT, but with realistic
\MET resolution, the \pTl distribution has essentially the same sensitivity to \mW. 
For example, in the CDF measurement~\cite{Aaltonen:2013vwa,Aaltonen:2012bp},
the statistical uncertainties with the muon channel are $16$~MeV and $18$~MeV for the \mT and \pTl fits, respectively.

Measurements are in progress by the ATLAS and CMS experiments at the LHC
and the high pileup environment means that they may focus 
on \pTl as their main fit variable
~\cite{Buge:951371,Besson:2008zs,ATL-PHYS-PUB-2014-015}.
At the LHC, the $W$ production cross section is roughly an order of magnitude higher than at the Tevatron,
so the statistical uncertainties will retire from their dominant status.
In the $\sqrt{s}=1.96$~TeV $p\bar{p}$ collisions at the Tevatron, 
$W$ production was dominated by valence $u\bar{d}$ and $d\bar{u}$ annihilation.
At the LHC, $W$ production receives a larger contribution from sea quarks.
Furthermore, the flavour composition is richer, with $\mathcal{O}(20\%)$ 
from $c\bar{s}/s\bar{c}$ annihilation.
A \mW measurement at the LHC is therefore subject to potentially limiting PDF uncertainties.
Ref.~\cite{Krasny:2010vd} casts a rather pessimistic outlook, while some
recent studies are more optimistic~\cite{Bozzi:2011ww,Rojo:2013nia,Bozzi:2015hha,Quackenbush:2015yra}.
Ref.~\cite{Bozzi:2015hha} estimates an uncertainty of around 20~MeV using current PDF sets.
However, this could be reduced to around 10~MeV with the requirement, $\pTW < 15$~GeV on the lepton-pair transverse momentum;
this cut makes the shape of the charged-lepton transverse momentum distribution steeper and closer to the leading-order one; it also suppresses the contribution from $qg$ scattering.
The studies reported in this paper assume that ATLAS and CMS will make this requirement in their measurements.
It is also pointed out in Ref.~\cite{Bozzi:2015hha} that the uncertainty would be greatly reduced
if the pseudorapidity, $\eta$, acceptance could be extended
from the roughly $|\eta| < 2.5$ of ATLAS and CMS, to $|\eta| < 4.9$,
because of an anti-correlation between the parton-parton luminosities and the lepton transverse momentum distribution at different charged lepton rapidities.
Ref.~\cite{Quackenbush:2015yra} proposed that even within the limited acceptance of the ATLAS and CMS
detectors, an exploitation of the correlations
between different rapidity regions and with the two $W$ charges could
reduce the uncertainty by around 60\%. 
Further improvements could be achieved by exploiting the correlations with \Zg decays~\cite{Quackenbush:2015yra}.

So far it has been assumed that ATLAS and CMS are the only LHC experiments with a chance
to improve on the direct \mW determination.
The LHCb experiment~\cite{LHCb-TDR-009} has not been discussed in this context.
\begin{itemize}
\item Firstly, the rate of $W$ production is far smaller in LHCb due to (i) the limited angular acceptance
$2 < \eta < 5$ and (ii) the lower instantaneous luminosity~\footnote{In 2012, LHCb already received a factor of twenty lower instantaneous luminosity than ATLAS and CMS, as required for the beauty and charm physics program.}.
\item Secondly, LHCb lacks the hermetic calorimeter coverage that is needed to reconstruct the \MET and \mT. The only available observable is thus \pTl.
\end{itemize}
In this paper, we argue that $W$ production in LHCb is sufficient to make a competitive measurement,
using the \pTl distribution,
and quantify the sensitivity with current and future datasets.
The unique angular acceptance of LHCb turns out to be a complement to the ATLAS and CMS measurements when we consider the PDF uncertainties.
In fact, the ability for LHCb to select a pure sample of $W \rightarrow \mu\nu$ decays 
without any requirement on the \MET, as already demonstrated in~\cite{Aaij:1439627}, 
is likely to be an advantage.
A key challenge in the Tevatron \mW measurements was the 
calibration of the detector response to the hadronic recoil.
This would be completely avoided in the LHCb measurement, which essentially
only requires knowledge of the muon reconstruction.~\footnote{It should be noted that current LHCb studies of $W$ production have imposed tight isolation requirements on the muon~\cite{Aaij:1439627}, 
the efficiency of which has some sensitivity to the hadronic recoil model.}
The present study has been performed assuming a given production model of the $W$ boson, i.e. making definite choices for the description of the QCD corrections that affect the \pTl distribution. 

In Sect.~\ref{sec:PDFs} the study of PDF uncertainties reported in Ref.~\cite{Bozzi:2015hha}
is extended to consider the impact of a LHCb measurement.
In Sect.~\ref{sec:LHCb}, the expected experimental uncertainties on a \mW measurement are estimated.
We choose to focus on the data that will be collected during Run-II (2015-2018) at a centre of mass
energy of $\sqrt{s} = 13$~TeV. 
It is expected that LHCb will record at least 7~fb$^{-1}$.
The prospects for a LHC combination are discussed in Sect.~\ref{sec:Prospects}.
In Sect.\ref{sec:ptwmodel} we comment on the uncertainties 
stemming from the \pTW modelling.

\section{\label{sec:PDFs}PDF uncertainties}

Our analysis is based on exactly the same setup as in Ref.~\cite{Bozzi:2015hha}.
Events of the type $pp \rightarrow W \rightarrow \ell\nu+X$, at $\sqrt{s}=13$~TeV,
are generated using POWHEG~\cite{Alioli:2008gx}, with parton showering provided by 
PYTHIA~\cite{Sjostrand:2006za}.
Replica templates for the \pTl distribution are produced for each of the 
NNPDF3.0~\cite{Ball:2014uwa}, MMHT2014~\cite{Harland-Lang:2014zoa} and CT10~\cite{CT10} PDF sets.
For the sake of simplicity, we assume the same kinematic acceptance for the ATLAS and CMS experiments,
and henceforth refer to them generically as the General Purpose Detector (GPD) experiments.
The GPD acceptance is defined as;
$|\eta| < 2.5$, $\pTl > 25$~GeV, $\pTnu > 25$~GeV, $\pTW < 15$~GeV.~\footnote{We assume that the GPD experiments will adopt the suggestion of Ref.~\cite{Bozzi:2015hha}, to require $\pTW < 15$~GeV.}
For LHCb, the kinematic acceptance is defined to be $2.0 < \eta < 4.5$ and  $\pTl > 20$~GeV.
The possibility of cut on $\pTnu$ and/or $\pTW$ is obviously excluded for LHCb.
For simplicity, we assume a  GPD averaged measurement for each $W$ charge,
already averaged over electron and muon channels.
In the following, these are denoted $\mathbf{G^+}$ and $\mathbf{G^-}$. 
The two LHCb measurements with $W \rightarrow \mu\nu$ are denoted $\mathbf{L^+}$ and $\mathbf{L^-}$.

\begin{figure}[b!]\centering
\includegraphics[width=0.49\linewidth]{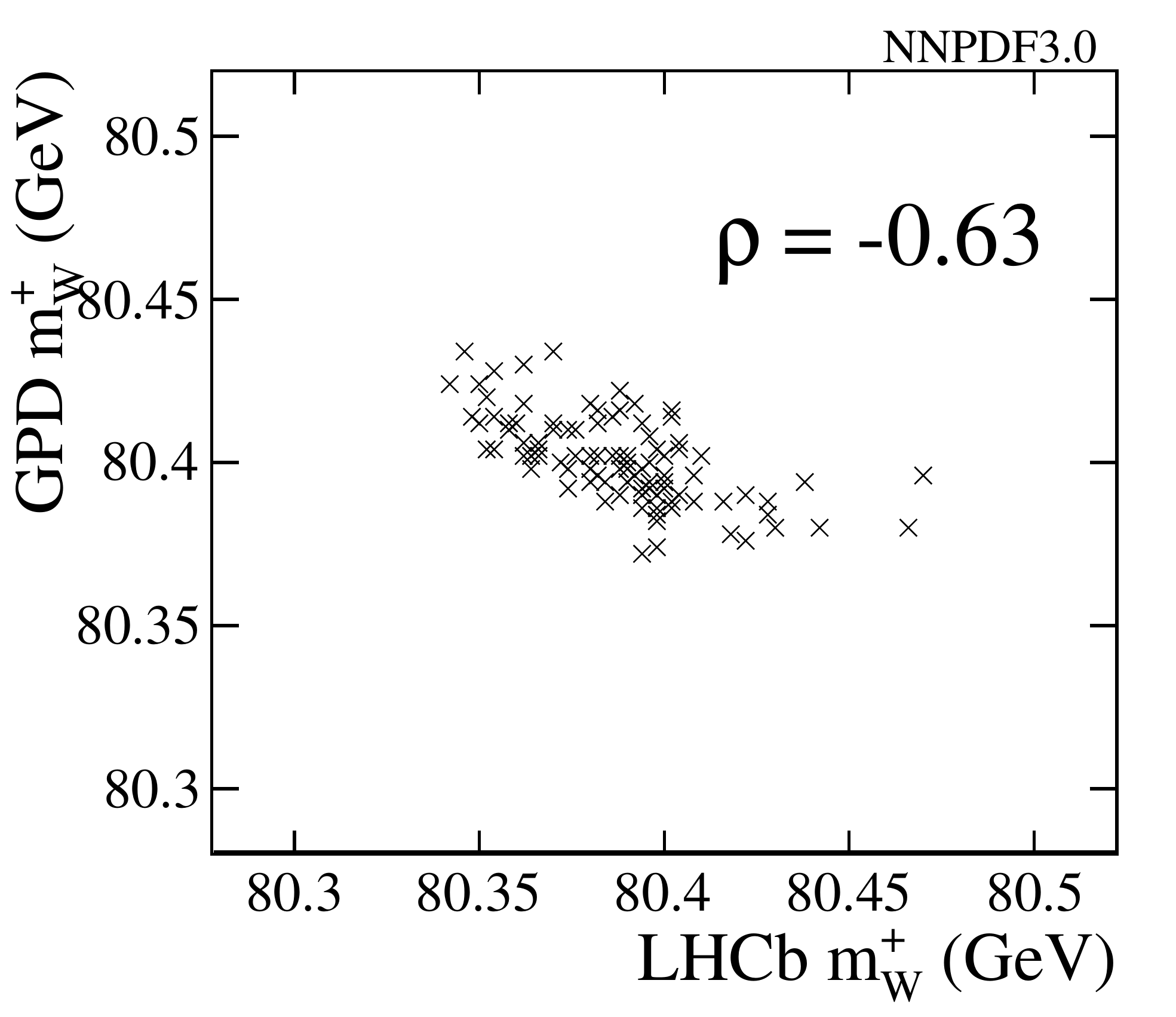}
\includegraphics[width=0.49\linewidth]{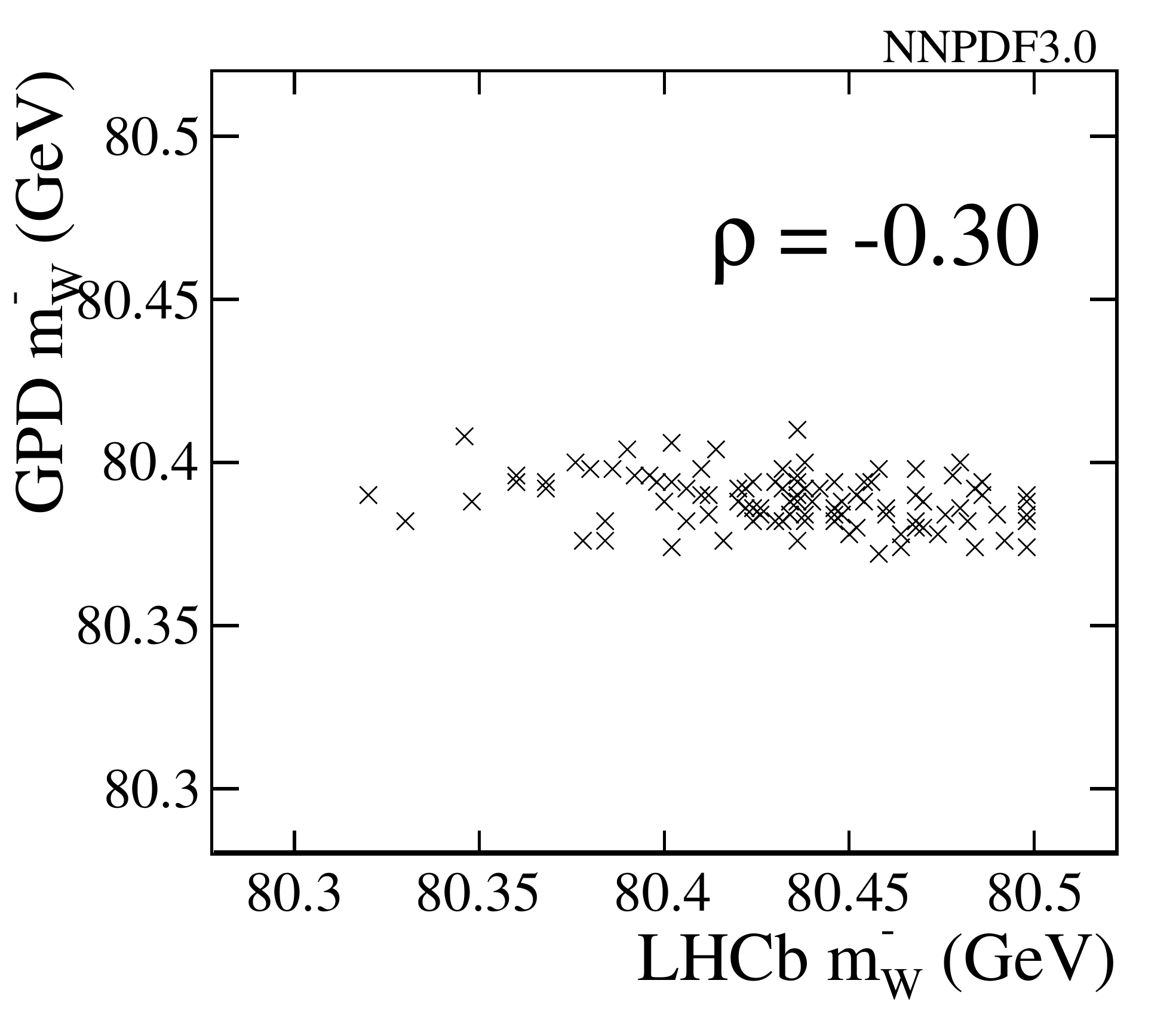}
\caption{\label{fig:mW_correlations_experiment}The fitted \mW\
in the GPDs versus LHCb for each NNPDF3.0 set, and for (left) $W^+$ and (right) $W^-$.}
\end{figure}

\begin{figure}[b!]\centering
\includegraphics[width=0.49\linewidth]{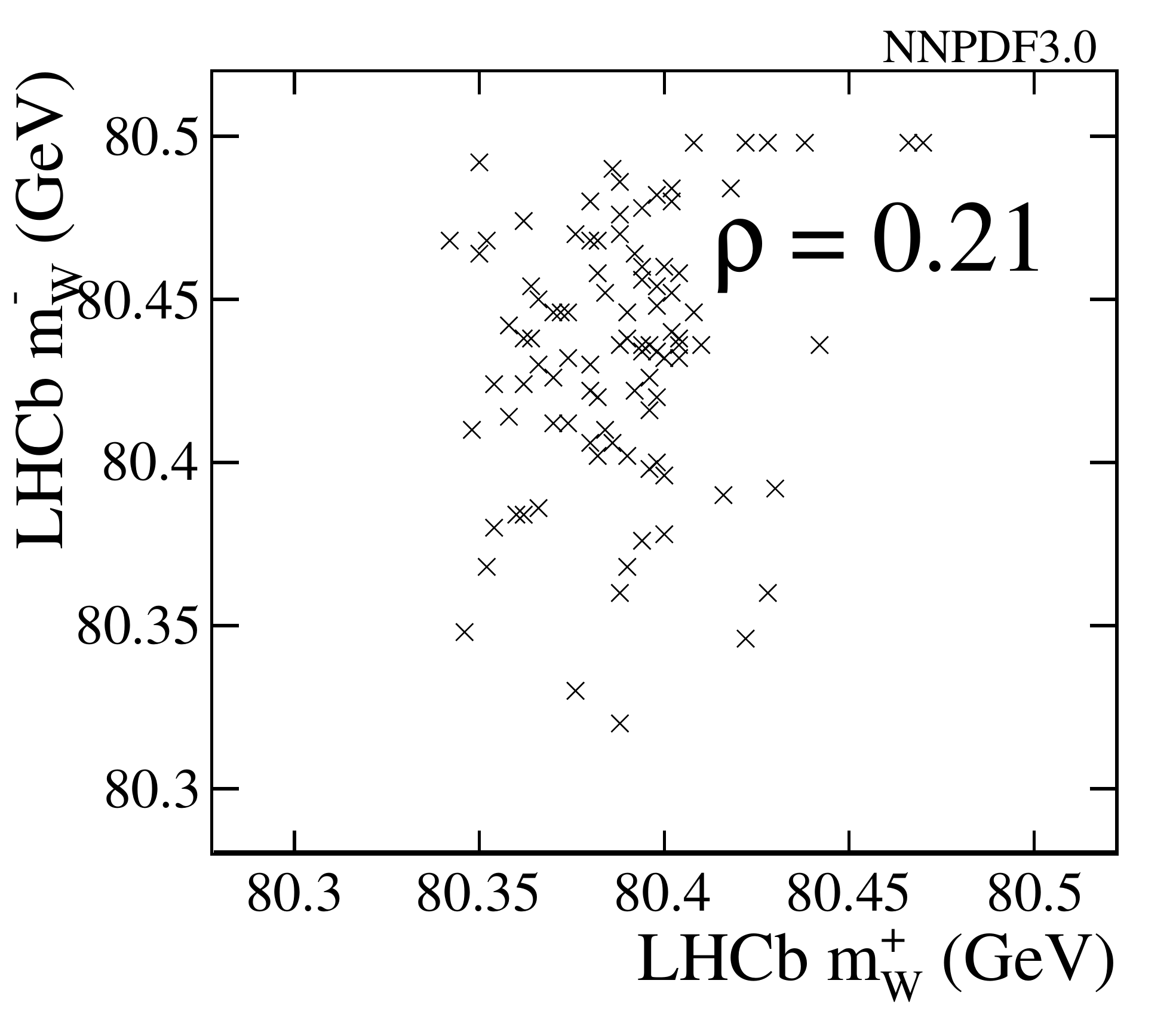}
\includegraphics[width=0.49\linewidth]{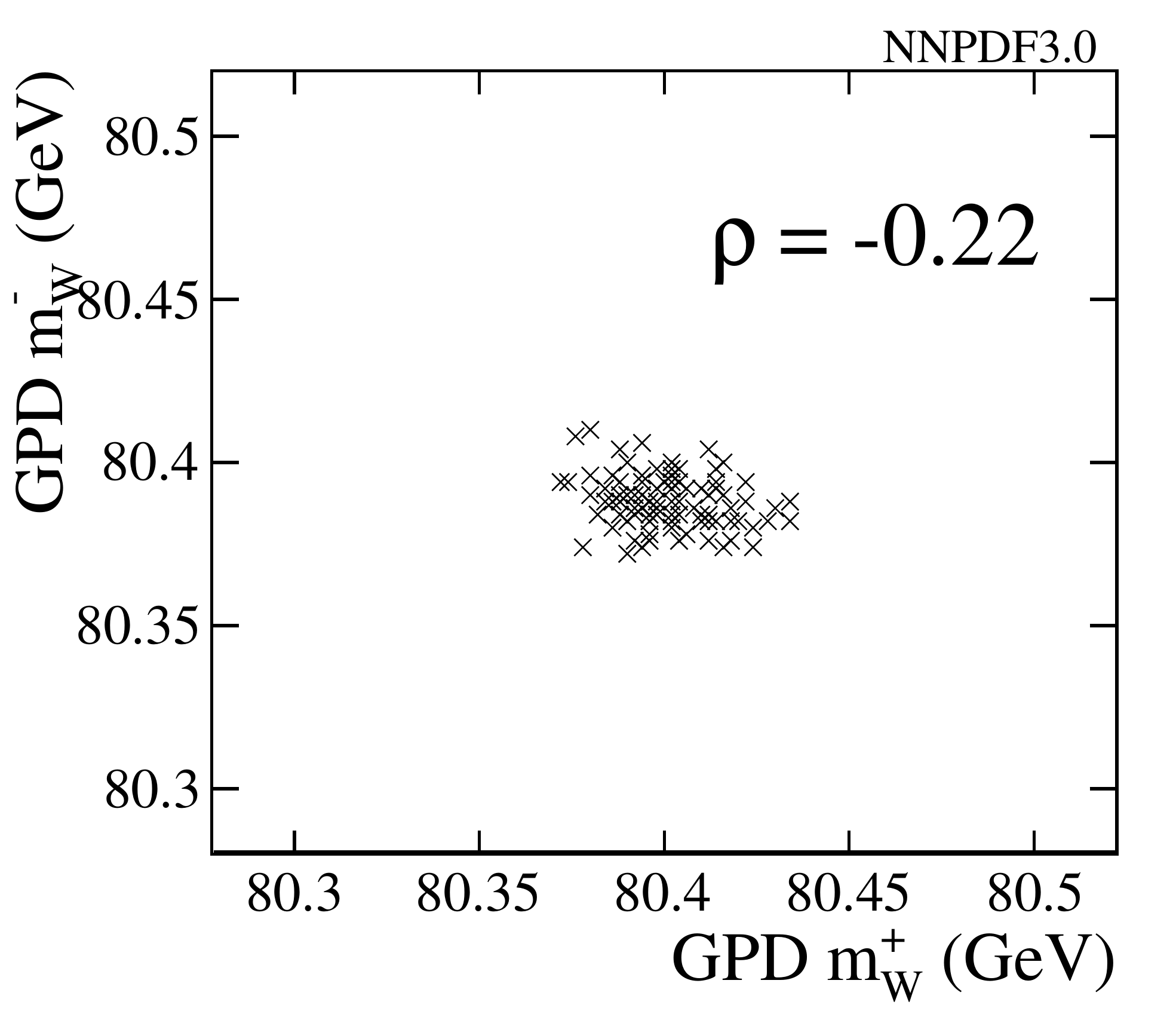}
\caption{\label{fig:mW_correlations_charge}The fitted \mW\
for $W^+$ versus $W^-$ and for (left) LHCb and (right) the GPDs.
Based in the NNPDF3.0 PDF sets.}
\end{figure}

We follow the PDF4LHC recommendation~\cite{Botje:2011sn} in estimating the 
PDF uncertainty.
If we consider the three sets (NNPDF3.0, MMHT2014, and CT10),
then the full uncertainty envelope of the considered sets is used.
In our default evaluation, we only consider the two most recent sets (NNPDF3.0 and MMHT2014),
which already include constraints from LHC data.
The following uncertainties
(in MeV) are estimated:
\begin{equation}
\deltaPDF = \makeatletter{}\left( \begin{array}{cc}\LabelGP & 24.8\\\LabelGM & 13.2\\\LabelLP & 27.0\\\LabelLM & 49.3\\\end{array} \right) ,
\end{equation}
These are repeated in Tab.~\ref{tab:PDF_errors_2sets}, while
Tab.~\ref{tab:PDF_errors_3sets} lists the corresponding uncertainties
that are evaluated with the inclusion of the CT10 sets.
In both tables, we also provide the largest difference in central values, denoted $\Delta_{\rm sets}$, 
between the (two or three) sets under consideration in each case.
This is evidently a major contributor to the uncertainty envelope.
For the $W^+$, similar uncertainties are estimated for LHCb and the GPDs.
For the $W^-$ on the other hand, the LHCb uncertainty is roughly a factor of four larger, because of the larger uncertainty of the sea quarks at large partonic $x$.
The real power of the LHCb measurement is revealed in the correlations.
With the NNPDF3.0 sets, we obtain the following correlation matrix:
\begin{equation}
\rho = \makeatletter{}\left( \begin{array}{ccccc}& \mathbf{G^{+}}& \mathbf{G^{-}}& \mathbf{L^{+}}& \mathbf{L^{-}}\\\mathbf{G^{+}} & 1&&&\\\mathbf{G^{-}} & -0.22&1&&\\\mathbf{L^{+}} & -0.63&0.11&1&\\\mathbf{L^{-}} & -0.02&-0.30&0.21&1\\\end{array} \right) .
\end{equation}
Similar correlation coefficients are found with the two other sets under consideration.
There is a particularly large negative correlation of around $-60$\% between the LHCb and GPD
measurements with the $W^+$, and a smaller anti-correlation of around $-30$\% for the $W^-$.
This can be clearly seen in Fig.~\ref{fig:mW_correlations_experiment} which
shows the distribution of fitted \mW values in the GPDs versus LHCb for the 100 NNPDF3.0 replicas.
For a single experiment, there are smaller correlations between the $W^+$ and $W^-$ measurements,
as can be seen in Fig.~\ref{fig:mW_correlations_charge}.
In LHCb, this is around $+20$\%, and for the GPDs, it is around $-20$\%.
Between different charges and experiments, the correlations are around 10\% or less in magnitude.
The normalised set of weights $\alpha_i$ that minimises the uncertainty on the weighted average of the four measurements $\mW_i$,
\begin{equation}
\mW = \sum\limits_{i=1}^4 \alpha_i \mW_i,
\end{equation}
would be
\begin{equation}
\alpha = \makeatletter{}\left( \begin{array}{cc}\mathbf{G+} & 0.30\\\mathbf{G-} & 0.45\\\mathbf{L+} & 0.21\\\mathbf{L-} & 0.04\\\end{array} \right) 
\end{equation}
The resulting PDF uncertainty would be $10.5$~MeV with the GPDs alone, and $7.7$~MeV including LHCb.
Tab.~\ref{tab:PDFErrors} lists the PDF uncertainties, with and without including LHCb.
The set of weights is also listed.
An average that includes \LabelLP with around 20\% of the weight, and with only a few percent 
for \LabelLM, would have a PDF uncertainty that is reduced by more than 30\%.
Tab.~\ref{tab:PDFErrors} also lists the corresponding numbers for scenarios in which:
\begin{itemize}
\item The CT10 sets are included in the uncertainty estimates:
In this case the PDF uncertainties are increased by roughly a factor of two,
but the relative impact of the LHCb measurement is similar to the 2-set scenario.
\item Each PDF set is considered separately:
instead of the envelope, the individual uncertainty bands are used.
The uncertainties are far smaller, but LHCb still has a large impact.
For the NNPDF3.0 sets, the gain is still around 30\%.
For the other two sets, the gain is closer to a factor of two!
\end{itemize}
The next question is whether or not LHCb can measure \mW with sufficient experimental precision to exploit this
anti-correlation in PDF uncertainties.

\begin{table}\centering
\caption{\label{tab:PDF_errors_2sets}PDF uncertainties on \mW (MeV) with the PDF4LHC prescription
using the NNPDF3.0 and MMHT2014 sets, for the 4 sub-measurements as defined in the text.}
\makeatletter{}\begin{tabular}{l|cccc}
\hline
 & \LabelGP & \LabelGM & \LabelLP & \LabelLM \\
\hline
Envelope & 24.8 & 13.2 & 27.0 & 49.3\\
$\Delta_{\rm sets}$ & 20.9 & 5.7 & 12.1 & 22.9\\
\hline
\end{tabular}
 
\end{table}

\begin{table}\centering
\caption{\label{tab:PDF_errors_3sets}PDF uncertainties on \mW (MeV) with the PDF4LHC prescription
using the NNPDF3.0, MMHT2014 and CT10 sets, for the 4 sub-measurements as defined in the text.}
\makeatletter{}\begin{tabular}{l|cccc}
\hline
 & \LabelGP & \LabelGM & \LabelLP & \LabelLM \\
\hline
Envelope & 29.9 & 23.5 & 35.0 & 84.1\\
$\Delta_{\rm sets}$ & 22.0 & 23.7 & 24.0 & 74.0\\
\hline
\end{tabular}
 
\end{table}

\begin{table}\centering
\caption{\label{tab:PDFErrors}The PDF uncertainties on the LHC averages
including and excluding LHCb, resulting from the weighted average with the 
optimal weights, $\alpha$.}
\makeatletter{}\begin{tabular}{l|l|c|l}
\hline\T\B
PDFs & Experiments & $\deltaPDF$~(MeV) & $\alpha$ \\
\hline\T\B
PDF4LHC(2-sets) & 2$\times$GPD & $10.5$ & $(0.26,0.74,0,0)$\\
PDF4LHC(2-sets) & 2$\times$GPD~+~LHCb & $7.7$ & $(0.30,0.45,0.21,0.04)$\\
\hline\T\B
PDF4LHC(3-sets) & 2$\times$GPD & $16.9$ & $(0.50,0.50,0,0)$\\
PDF4LHC(3-sets) & 2$\times$GPD~+~LHCb & $12.7$ & $(0.43,0.41,0.11,0.04)$\\
\hline\T\B
NNPDF30 & 2$\times$GPD & $5.2$ & $(0.50,0.50,0,0)$\\
NNPDF30 & 2$\times$GPD~+~LHCb & $3.6$ & $(0.35,0.47,0.16,0.02)$\\
\hline\T\B
MMHT2014 & 2$\times$GPD & $9.2$ & $(0.45,0.55,0,0)$\\
MMHT2014 & 2$\times$GPD~+~LHCb & $4.6$ & $(0.39,0.14,0.46,0)$\\
\hline\T\B
CT10 & 2$\times$GPD & $11.6$ & $(0.33,0.67,0,0)$\\
CT10 & 2$\times$GPD~+~LHCb & $6.3$ & $(0.38,0.20,0.40,0.03)$\\
\hline
\end{tabular}
 
\end{table}

\section{\label{sec:LHCb}LHCb experimental sensitivity to the W mass}

In Run-I (2010-2012), LHCb recorded 3~fb$^{-1}$ of $pp$ collisions at $\sqrt{s}=7-8$~TeV.
In Run-II (2015-2018),  LHCb hopes to record around 7~fb$^{-1}$ at $\sqrt{s}=13$~TeV.
Given the $W \rightarrow \mu\nu$ signal yields reported in a LHCb measurement using only 1~fb$^{-1}$ 
of data from the 2011 Run~\cite{Aaij:2015gna}, we extrapolate the projected Run-I and Run-II signal yields,
and use these to estimate the uncertainties on a \mW measurement with LHCb.
These estimates are listed in Tab.~\ref{tab:exp_errors}, for both the Run-I and Run-II datasets.
They are quoted separately for the $W^+$ and $W^-$ since the PDF uncertainties,
as discussed in detail in Sect.~\ref{sec:PDFs}, motivate separate analyses for the two charges.

\begin{table}[!b]\centering
\begin{tabular}{r|cccc}
       & \multicolumn{2}{c}{Run-I} & \multicolumn{2}{c}{Run-II} \\
       & \multicolumn{2}{c}{3~fb$^{-1}$} & \multicolumn{2}{c}{7~fb$^{-1}$} \\
\hline
       & $W^+$ & $W^-$ & $W^+$ & $W^-$ \\
Signal yields, $\times 10^{6}$ & 1.2 & 0.7 & 5.4 & 3.4 \\
$Z/\gamma^*$ background, ($B/S$) & 0.15 & 0.15 & 0.15 & 0.15 \\ 
QCD background, ($B/S$) & 0.15 & 0.15 & 0.15 & 0.15 \\ 
$\delta_{\mW}$ (MeV) & & & &\\
Statistical & 19 & 29 & 9 & 12 \\
Momentum scale & 7 & 7 & 4 & 4 \\
Quadrature sum & 20 & 30 & 10 & 13\\
\end{tabular}
\caption{\label{tab:exp_errors}The estimated experimental uncertainties on a \mW measurement with LHCb.}
\end{table}

\subsection[Signal statistics]{Statistical sensitivity estimate for the \pTl fit}
\label{sec:StatErr}
In Ref.~\cite{Aaij:2015gna}, LHCb found, in 1~fb$^{-1}$ of Run-I data, 
around 550k candidate muonic $W^+$ decays, and around 350k $W^-$, with a purity of around 70\%.
The extrapolated signal yields in the full Run-I and Run-II datasets are listed in Tab.~\ref{tab:exp_errors}.
It is assumed that the cross sections for $W^{\pm}$ production increase by a factor of two 
with the increased collision energy in Run-II.
The Run-I yield of around two million can be compared with the 
0.6(0.5) million $W \rightarrow \mu(e)\nu$ candidates 
that were used in the CDF measurement with 2.1~fb$^{-1}$~\cite{Aaltonen:2013vwa,Aaltonen:2012bp}.
The D0 measurement with 4.3 fb$^{-1}$~\cite{D0:2013jba,Abazov:2012bv} used around
1.7 million $W \rightarrow e\nu$ signal candidates.
The Run-II $W \rightarrow \mu\nu$ yield in LHCb, assuming an integrated luminosity of 7~fb$^{-1}$, will be around eight million.

In order to estimate the statistical precision on the \mW fit with LHCb data,
we take the \pTl templates described in Sect.~\ref{sec:PDFs}.
The dominant background reported in Ref~\cite{Aaij:2015gna}
is $Z/\gamma^* \rightarrow \mu\mu$ where one muon escapes the limited angular acceptance of LHCb.
At lower \pTl, there is a large ``QCD'' background from muonic decays of pions and kaons.
Directly under the upper edge Jacobian peak, where most of the \mW sensitivity is delivered,
the QCD background is small.
An exponential parameterisation for each of the $\Zg$ and QCD backgrounds is added to the signal 
\pTl templates, with yields and shapes roughly reproducing those in Ref.~\cite{Aaij:2015gna}.
The signal and background templates are scaled to the projected yields listed in Tab.~\ref{tab:exp_errors}.
From this spectrum, we generate 500 copies but with each bin 
varied according to a Poisson random number.
Each of the 500 pseudo-datasets is compared to the ensemble of templates corresponding
to different \mW values.
The best fit value for each of these 500 copies is obtained from the minimum $\chi^2$ and with the uncertainty defined
by $\Delta\chi^2 = \pm 1$.
Tab.~\ref{tab:exp_errors} lists the statistical uncertainty computed as the spread of the best fit central values.~\footnote{
The $\pm 29$~MeV uncertainty for the $W^-$ in Run-I can be compared to the $\pm 22$~MeV
that was reported by CDF in the \pTl fit 
using a similar number of $W\rightarrow\mu\nu$ events~\cite{Aaltonen:2013vwa,Aaltonen:2012bp}.}
With the Run-II dataset, LHCb could achieve statistical uncertainties
of 10(13)~MeV for the $W^+$($W^-$).

\begin{figure}[b!]\centering
\includegraphics[width=0.49\linewidth]{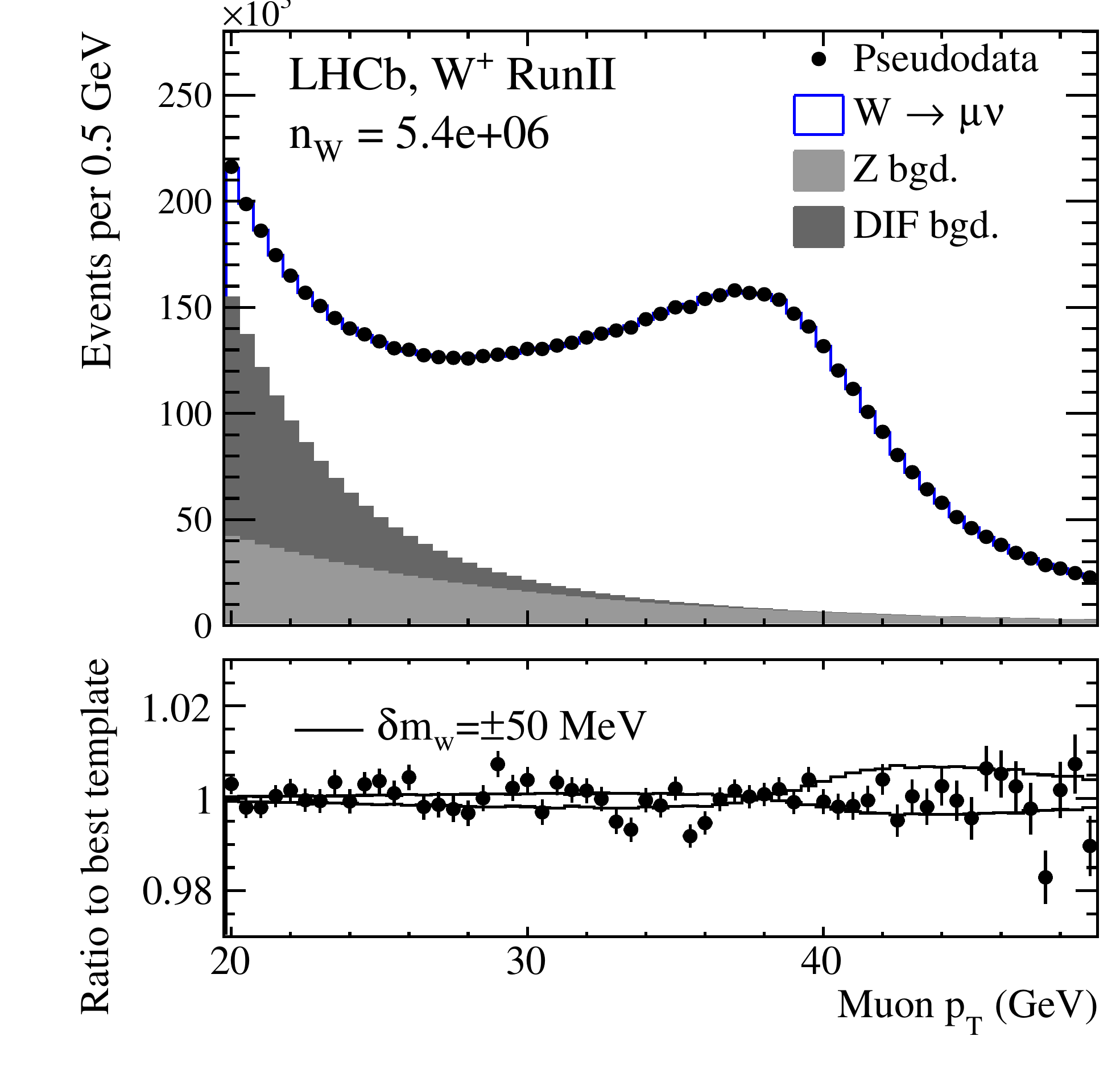}
\includegraphics[width=0.49\linewidth]{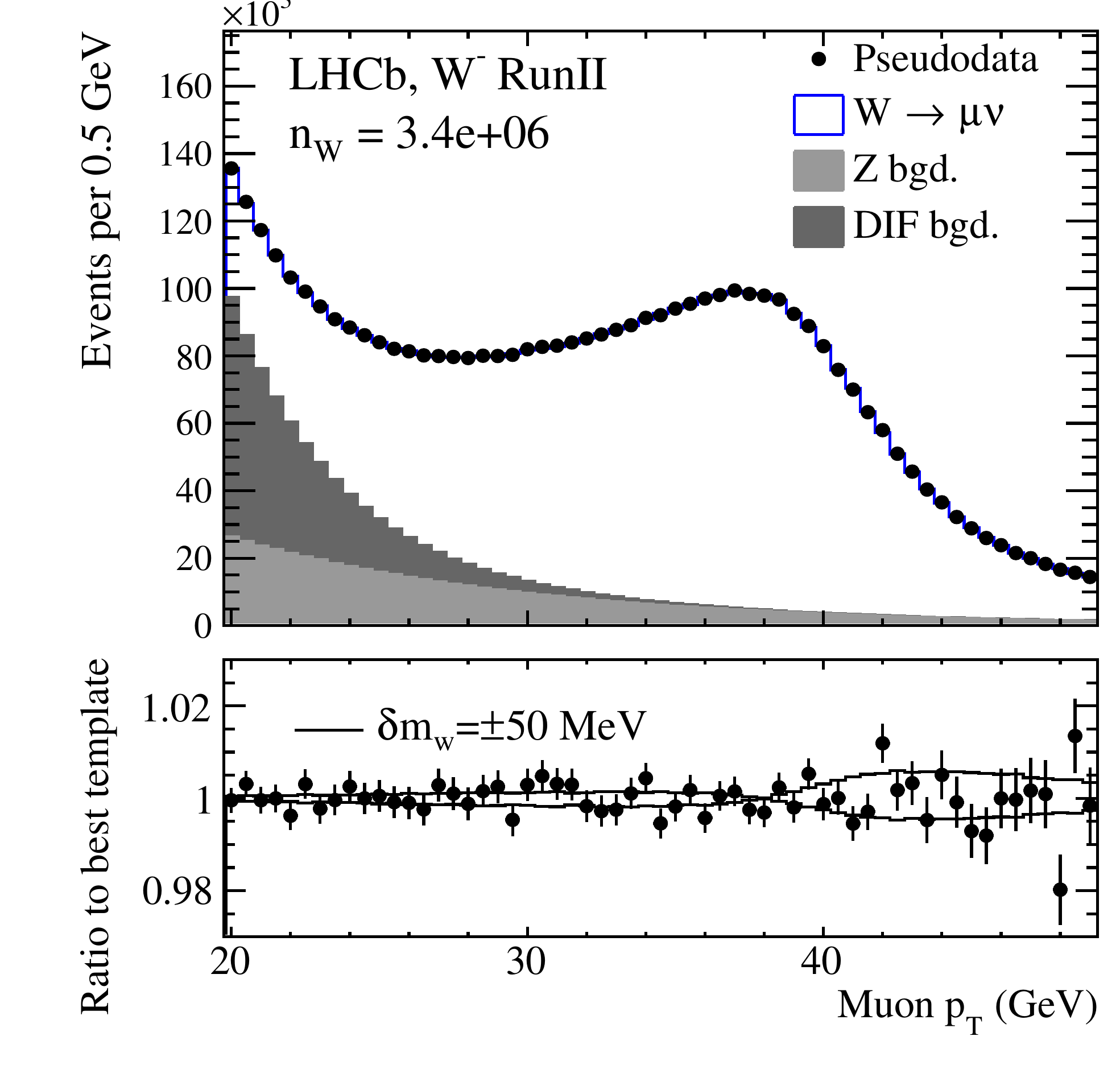}
\caption{\label{fig:toys_RunII}The simulated \pTl spectra for (left) $W^+$ and (right) $W^-$
decays in the projected Run-II LHCb dataset.
The data points correspond to one of the 500 pseudo-datasets.
The stacked histogram corresponds to the best fit template.
In the lower panel, the points represent the ratio of the data to the best fit template,
and the lines show the ratio of the best fit template to templates with \mW varying by $\pm 50$~MeV.}
\end{figure}

\subsection{Momentum scale calibration}
\label{sec:ScaleErr}
In the Tevatron \mW measurements, the muon momentum scale, and electron energy scale,
were major contributors to the total uncertainties on \mW.
In particular, the D0 measurement with $W \rightarrow e\nu$ relied almost entirely on $Z/\gamma^* \rightarrow e^+e^-$ events, leading to the single largest source of uncertainty of around 20~MeV, 
depending on the fit variable.
The CDF measurement exploited a combination of $J/\psi$, $\psi(2S)$, $\Upsilon(nS)$ ($n=1,2,3$) and $Z/\gamma^*$ decays into  $\mu^+\mu^-$ to achieve a muon momentum scale uncertainty of 7~MeV.

LHCb is ideally suited for a precise calibration of the momentum scale,
due to the large samples of inclusive quarkonia signals that are recorded.
Furthermore, LHCb has an excellent momentum resolution that ranges between 0.2\% and 0.8\%~\cite{LHCb-TDR-009}.
LHCb has already demonstrated its ability to make world's best measurements of various
$b$- and $c$-hadron masses~\cite{Aaij:2013qja,Aaij:2013uaa}.
In Ref.~\cite{Aaij:2013qja}, LHCb reported a relative momentum scale uncertainty of 
$3 \times 10^{-4}$ as part of a measurement of $b$-baryon masses, using only 35 pb$^{-1}$ of data.

The \ZtoMuMu line shape provides an important 
constraint on the momentum scale at high \pT.
Roughly speaking this would be $\delta_z/\sqrt{N}$, where $N$ is the number
of \Zg events in the $Z$ peak region, and the observed line-width, $\delta_z \sim 3$~GeV,
is dominated by the natural width of the $Z$.
A concern might be that while LHCb records enough $W$ decays,
the limited angular acceptance might not allow sufficient $\Zg$ samples.
In Ref.~\cite{Aaij:2015gna}, LHCb found, in 1~fb$^{-1}$ of Run-I data, 
around 60k $Z/\gamma^* \rightarrow \mu\mu$ candidates with almost perfect purity.
We estimate around 150k signal in the full Run-I dataset and around 700k in Run-II,
yielding naive momentum scale uncertainties of $7$~MeV and $3$~MeV, respectively.
Further constraints can be obtained from the $J/\psi$ and $\Upsilon$ resonances.
Extrapolating the $\Upsilon(1S)$ yields reported in~\cite{Aaij:1428520},
the full Run-I dataset should already provide a few million decays.
A dedicated study would be needed to demonstrate that the alignment of the LHCb 
tracking detectors could be understood with sufficient precision to relate these mass
constraints to the momentum scale. 
For the purpose of our present study, 
we assume a momentum scale uncertainty on \mW of 7(4) MeV for the Run-I(II) datasets.

\subsection{Muon efficiencies}
\label{sec:efficiencies}
The methods to measure muon reconstruction efficiencies in LHCb are well developed
as part of the $Z/\gamma^*$ cross section measurements~\cite{Aaij:1439627}.
Since the \mW measurement is only concerned with the kinematic dependence of the efficiency
and not in its absolute scale, it can safely be assumed that this will be a sub-dominant source
of uncertainty.

\section{\label{sec:Prospects}Prospects for an LHC \boldmath{$\mW$} combination}

The experimental precision with which ATLAS and CMS can measure \mW will no
doubt have evolved
since the discussions in Refs.~\cite{Buge:951371,Besson:2008zs}.
The idea of this study is not to make a precise estimate of the LHC
sensitivity, but rather to estimate the relative
impact of the proposed LHCb measurement.
Our assumption is that ATLAS and CMS can both measure \mW with experimental
uncertainties of $\pm$7 MeV 
per $W$ charge, having averaged over electron and muon decay channels.
Large variations either side of this assumption are considered in our study.

The four measurements would have the following uncertainties, using the NNPDF3.0 and MMHT2014 PDF sets,
\begin{equation}
\delta m_W^i = 
\left(
\begin{array}{cl}
\mathbf{G^+} &(7_{\rm exp} \pm 25_{\rm PDF})~\mathrm{MeV}\\
\mathbf{G^-} &(7_{\rm exp} \pm 13_{\rm PDF})~\mathrm{MeV}\\
\mathbf{L^+} &(10_{\rm exp} \pm 28_{\rm PDF})~\mathrm{MeV}\\
\mathbf{L^-} &(13_{\rm exp} \pm 49_{\rm PDF})~\mathrm{MeV}\\
\end{array}
\right).
\end{equation}
For the sake of simplicity, our study only considers experimental and PDF errors, while for a more realistic estimate one should include also 
other sources of theoretical uncertainty.
For each experiment, we assume a positive correlation of $50$\% between the
experimental uncertainties for $W^+$ and $W^-$, as can be expected since 
many experimental calibrations are independent of the charge.
The set of weights are optimised to give the smallest total uncertainty
on the weighted average of the four measurements.
The resulting uncertainties and optimal weights are listed in Tab.~\ref{tab:TotalErrors}.
The first three rows show the results of 
(i) the LHC average including all three experiments with muons and electrons (only for the GPDs) and both charges,
(ii) a combination of LHCb and one GPD,
(iii) a combination of the two GPDs without LHCb.
The total uncertainty is improved by around 30\% when LHCb is included.
Interestingly, an average of LHCb with a single GPD would be more precise than a two-GPD combination.
Tab.~\ref{tab:TotalErrors} also lists the 
corresponding uncertainties and weights for a number of variations in our assumptions.
\begin{itemize}
\item If all three PDF sets are used with the PDF4LHC prescription, 
the total uncertainty is larger, and the impact of LHCb is even larger than with the two more recent sets.
\item We consider the four possibilities of setting the LHCb or GPD experimental uncertainties to zero or twice our nominal assumption.
In all cases, LHCb is more important in the average, than a second GPD.
\item Not surprisingly, LHCb has more(less) impact if we scale the PDF uncertainties by a factor of two(zero).
\end{itemize}
It seems that in any realistic scenario, excluding the extreme cases above,
LHCb would reduce the total uncertainty on the LHC average by 20--40\%.
And in all of these scenarios, we have the remarkable result that LHCb has more impact
than a second GPD.

\begin{table}\centering
\caption{\label{tab:TotalErrors}The uncertainties on different LHC averages for \mW.
The separate experimental and PDF uncertainties are listed, as are the weights
that minimise the total uncertainty.}
\makeatletter{}\begin{tabular}{r|l|c|c|c|l}
\hline
          &             & \multicolumn{3}{c|}{$\delta m_W$~(MeV)} & \\
 Scenario & Experiments & Tot & Exp & PDF                           & $\alpha$ \\
\hline \T\B
Default & 2$\times$GPD~+~LHCb & 9.0 & 4.7 & 7.7 & $(0.30,0.44,0.22,0.04)$\\
Default & 1$\times$GPD~+~LHCb & 10.1 & 6.5 & 7.7 & $(0.31,0.40,0.25,0.04)$\\
Default & 2$\times$GPD & 12.0 & 5.8 & 10.5 & $(0.28,0.72,0,0)$\\
\hline \T\B
PDF4LHC(3-sets) & 2$\times$GPD~+~LHCb & 13.6 & 4.8 & 12.7 & $(0.43,0.41,0.12,0.04)$\\
PDF4LHC(3-sets) & 1$\times$GPD~+~LHCb & 14.6 & 7.3 & 12.7 & $(0.43,0.40,0.12,0.04)$\\
PDF4LHC(3-sets) & 2$\times$GPD & 17.7 & 5.5 & 16.9 & $(0.50,0.50,0,0)$\\
\hline \T\B
$\delta_{\rm exp}^{\rm LHCb} = 0$ & 2$\times$GPD~+~LHCb & 8.7 & 4.0 & 7.7 & $(0.31,0.41,0.24,0.04)$\\
$\delta_{\rm exp}^{\rm LHCb} = 0$ & 1$\times$GPD~+~LHCb & 9.8 & 5.9 & 7.9 & $(0.31,0.37,0.28,0.04)$\\
$\delta_{\rm exp}^{\rm LHCb} = 0$ & 2$\times$GPD & 12.0 & 5.8 & 10.5 & $(0.28,0.72,0,0)$\\
\hline \T\B
$\delta_{\rm exp}^{\rm GPD} = 0$ & 2$\times$GPD~+~LHCb & 7.9 & 1.9 & 7.7 & $(0.29,0.48,0.19,0.04)$\\
$\delta_{\rm exp}^{\rm GPD} = 0$ & 1$\times$GPD~+~LHCb & 7.9 & 1.9 & 7.7 & $(0.29,0.48,0.19,0.04)$\\
$\delta_{\rm exp}^{\rm GPD} = 0$ & 2$\times$GPD & 10.5 & 0.1 & 10.5 & $(0.26,0.74,0,0)$\\
\hline \T\B
$\delta_{\rm PDF} = 0$ & 2$\times$GPD~+~LHCb & 4.6 & 4.6 & 0.0 & $(0.34,0.34,0.22,0.10)$\\
$\delta_{\rm PDF} = 0$ & 1$\times$GPD~+~LHCb & 5.8 & 5.8 & 0.0 & $(0.23,0.23,0.37,0.17)$\\
$\delta_{\rm PDF} = 0$ & 2$\times$GPD & 5.5 & 5.5 & 0.0 & $(0.50,0.50,0,0)$\\
\hline \T\B
$\delta_{\rm exp}^{\rm LHCb} \times 2$ & 2$\times$GPD~+~LHCb & 9.6 & 5.6 & 7.7 & $(0.29,0.50,0.17,0.04)$\\
$\delta_{\rm exp}^{\rm LHCb} \times 2$ & 1$\times$GPD~+~LHCb & 10.8 & 7.6 & 7.7 & $(0.30,0.46,0.20,0.05)$\\
$\delta_{\rm exp}^{\rm LHCb} \times 2$ & 2$\times$GPD & 12.0 & 5.8 & 10.5 & $(0.28,0.72,0,0)$\\
\hline \T\B
$\delta_{\rm exp}^{\rm GPD} \times 2$ & 2$\times$GPD~+~LHCb & 11.2 & 7.9 & 8.0 & $(0.32,0.35,0.29,0.04)$\\
$\delta_{\rm exp}^{\rm GPD} \times 2$ & 1$\times$GPD~+~LHCb & 13.9 & 10.5 & 9.0 & $(0.31,0.26,0.37,0.05)$\\
$\delta_{\rm exp}^{\rm GPD} \times 2$ & 2$\times$GPD & 15.6 & 11.5 & 10.6 & $(0.32,0.68,0,0)$\\
\hline \T\B
$\delta_{\rm PDF} \times 2$ & 2$\times$GPD~+~LHCb & 16.0 & 4.7 & 15.3 & $(0.30,0.45,0.21,0.04)$\\
$\delta_{\rm PDF} \times 2$ & 1$\times$GPD~+~LHCb & 16.7 & 6.7 & 15.3 & $(0.30,0.44,0.22,0.04)$\\
$\delta_{\rm PDF} \times 2$ & 2$\times$GPD & 21.7 & 5.9 & 20.9 & $(0.27,0.73,0,0)$\\
\hline
\end{tabular}
 
\end{table}

\section{Uncertainties stemming from the \pTW modelling}
\label{sec:ptwmodel}
Another source of theoretical uncertainty that we have overlooked so far is the \pTW model. This strongly affects the preparation of the templates that are used to fit the data and eventually to extract \mW.
The presence, at low lepton-pair transverse momenta, of large logarithmically enhanced QCD corrections and the role, in the same kinematic region, of non-perturbative effects have been discussed in Refs.~\cite{Landry:2002ix,Konychev:2005iy}, but the dependence of the resulting model on the acceptance cuts has never been investigated in detail and will deserve a dedicated study.
The \pTl is more sensitive to this than \mT . 
At the Tevatron, the uncertainty from the \pTW model on the \pTl fit was around 5 MeV, but perturbative QCD scale uncertainties should also be taken into account.
To a first approximation the results of the present note are independent of the uncertainty stemming from on the \pTW modelling and will hopefully be confirmed if the latter will turn out to be under control.
On a longer term perspective we will need a global analysis of all the non-perturbative elements active in the proton description: the PDFs uncertainties, in particular the role of heavy quarks in the proton~\cite{Berge:2005rv,ATL-PHYS-PUB-2014-015}, and the description of the intrinsic transverse momentum of the partons. The inclusion of all the different Drell-Yan channels (neutral current, $W^+$ and $W^-$) in the different acceptance regions of the LHC experiments might have an impact on a systematic reduction of all these uncertainties.

\section{Summary}
Improving the precision on \mW remains a priority in particle physics.
At the LHC, there is no shortage of $W$ production, but there are potentially limiting PDF uncertainties
on the anticipated measurements by ATLAS and CMS, which cover central lepton pseudorapidities, $|\eta| \lesssim 2.5$.
We show that a measurement in the forward acceptance of the LHCb experiment, $2 < \eta < 4.5$, 
would have a PDF uncertainty that is highly anti-correlated with those of ATLAS and CMS.
In this paper we study the measurement of \mW extracted from the \pTl distribution.
The weighted average of the ATLAS, CMS and LHCb results, based only on the PDF uncertainties,
would be 30\% more precise than an average of ATLAS and CMS alone.
Despite the lower rate of $W$ production in LHCb, a measurement could be made with the Run-II dataset,
using $W \rightarrow \mu\nu$ decays, that improves the total uncertainty on the LHC average by 20-40\%, depending on the assumptions on the experimental uncertainties. 
In fact, for any realistic scenario, LHCb has more impact in the LHC average than a second GPD.
It remains to be demonstrated that the \pTW model uncertainties can be controlled at the necessary level of precision, 
but deeper study into a possible \mW measurement with LHCb, and its combination with 
the ATLAS and CMS measurements, is well motivated.

\begin{acknowledgements}
We are thank full to the following members of the LHCb Collaboration for 
helpful discussions: Will Barter, Simone Bifani, Victor Coco, Stephen Farry, Wouter Hulsbergen, Phil Ilten, Patrick Koppenburg, Tara Shears. 
M.~V.~is supported by the Alexander von Humboldt Foundation.
A.~V.~is supported in part by an Italian PRIN2010 grant, by a European Investment Bank EIBURS grant, and by the European Commission through the HiggsTools Initial Training Network PITN-GA-2012-316704. 
\end{acknowledgements}

\bibliographystyle{LHCb}
\bibliography{main} 

\ifx\mcitethebibliography\mciteundefinedmacro
\PackageError{LHCb.bst}{mciteplus.sty has not been loaded}
{This bibstyle requires the use of the mciteplus package.}\fi
\providecommand{\href}[2]{#2}
\begin{mcitethebibliography}{10}
\mciteSetBstSublistMode{n}
\mciteSetBstMaxWidthForm{subitem}{\alph{mcitesubitemcount})}
\mciteSetBstSublistLabelBeginEnd{\mcitemaxwidthsubitemform\space}
{\relax}{\relax}

\bibitem{Baak:2014ora}
Gfitter Group, M.~Baak {\em et~al.},
  \ifthenelse{\boolean{articletitles}}{\emph{{The global electroweak fit at
  NNLO and prospects for the LHC and ILC}},
  }{}\href{http://dx.doi.org/10.1140/epjc/s10052-014-3046-5}{Eur.\ Phys.\ J.\
  \textbf{C74} (2014) 3046}, \href{http://arxiv.org/abs/1407.3792}{{\tt
  arXiv:1407.3792}}\relax
\mciteBstWouldAddEndPuncttrue
\mciteSetBstMidEndSepPunct{\mcitedefaultmidpunct}
{\mcitedefaultendpunct}{\mcitedefaultseppunct}\relax
\EndOfBibitem
\bibitem{Awramik:2003rn}
M.~Awramik, M.~Czakon, A.~Freitas, and G.~Weiglein,
  \ifthenelse{\boolean{articletitles}}{\emph{{Precise prediction for the W
  boson mass in the standard model}},
  }{}\href{http://dx.doi.org/10.1103/PhysRevD.69.053006}{Phys.\ Rev.\
  \textbf{D69} (2004) 053006}, \href{http://arxiv.org/abs/hep-ph/0311148}{{\tt
  arXiv:hep-ph/0311148}}\relax
\mciteBstWouldAddEndPuncttrue
\mciteSetBstMidEndSepPunct{\mcitedefaultmidpunct}
{\mcitedefaultendpunct}{\mcitedefaultseppunct}\relax
\EndOfBibitem
\bibitem{Degrassi:2014sxa}
G.~Degrassi, P.~Gambino, and P.~P. Giardino,
  \ifthenelse{\boolean{articletitles}}{\emph{{The $m_{\scriptscriptstyle
  W}-m_{\scriptscriptstyle Z}$ interdependence in the Standard Model: a new
  scrutiny}}, }{}\href{http://dx.doi.org/10.1007/JHEP05(2015)154}{JHEP
  \textbf{05} (2015) 154}, \href{http://arxiv.org/abs/1411.7040}{{\tt
  arXiv:1411.7040}}\relax
\mciteBstWouldAddEndPuncttrue
\mciteSetBstMidEndSepPunct{\mcitedefaultmidpunct}
{\mcitedefaultendpunct}{\mcitedefaultseppunct}\relax
\EndOfBibitem
\bibitem{PDG2014}
Particle Data Group, K.~A. Olive {\em et~al.},
  \ifthenelse{\boolean{articletitles}}{\emph{{\href{http://pdg.lbl.gov/}{Review
  of particle physics}}},
  }{}\href{http://dx.doi.org/10.1088/1674-1137/38/9/090001}{Chin.\ Phys.\
  \textbf{C38} (2014) 090001}\relax
\mciteBstWouldAddEndPuncttrue
\mciteSetBstMidEndSepPunct{\mcitedefaultmidpunct}
{\mcitedefaultendpunct}{\mcitedefaultseppunct}\relax
\EndOfBibitem
\bibitem{Heinemeyer:2006px}
S.~Heinemeyer {\em et~al.}, \ifthenelse{\boolean{articletitles}}{\emph{{Precise
  prediction for M(W) in the MSSM}},
  }{}\href{http://dx.doi.org/10.1088/1126-6708/2006/08/052}{JHEP \textbf{0608}
  (2006) 052}, \href{http://arxiv.org/abs/hep-ph/0604147}{{\tt
  arXiv:hep-ph/0604147}}\relax
\mciteBstWouldAddEndPuncttrue
\mciteSetBstMidEndSepPunct{\mcitedefaultmidpunct}
{\mcitedefaultendpunct}{\mcitedefaultseppunct}\relax
\EndOfBibitem
\bibitem{Aaltonen:2013vwa}
T.~A. Aaltonen {\em et~al.},
  \ifthenelse{\boolean{articletitles}}{\emph{{Precise measurement of the W
  -boson mass with the Collider Detector at Fermilab}},
  }{}\href{http://dx.doi.org/10.1103/PhysRevD.89.072003}{Phys.\ Rev.\
  \textbf{D89} (2014), no.~7 072003},
  \href{http://arxiv.org/abs/1311.0894}{{\tt arXiv:1311.0894}}\relax
\mciteBstWouldAddEndPuncttrue
\mciteSetBstMidEndSepPunct{\mcitedefaultmidpunct}
{\mcitedefaultendpunct}{\mcitedefaultseppunct}\relax
\EndOfBibitem
\bibitem{Aaltonen:2012bp}
T.~Aaltonen {\em et~al.}, \ifthenelse{\boolean{articletitles}}{\emph{{Precise
  measurement of the $W$-boson mass with the CDF II detector}},
  }{}\href{http://dx.doi.org/10.1103/PhysRevLett.108.151803}{Phys.\ Rev.\
  Lett.\  \textbf{108} (2012) 151803},
  \href{http://arxiv.org/abs/1203.0275}{{\tt arXiv:1203.0275}}\relax
\mciteBstWouldAddEndPuncttrue
\mciteSetBstMidEndSepPunct{\mcitedefaultmidpunct}
{\mcitedefaultendpunct}{\mcitedefaultseppunct}\relax
\EndOfBibitem
\bibitem{D0:2013jba}
D0, V.~M. Abazov {\em et~al.},
  \ifthenelse{\boolean{articletitles}}{\emph{{Measurement of the $W$ boson mass
  with the D0 detector}},
  }{}\href{http://dx.doi.org/10.1103/PhysRevD.89.012005}{Phys.\ Rev.\
  \textbf{D89} (2014), no.~1 012005},
  \href{http://arxiv.org/abs/1310.8628}{{\tt arXiv:1310.8628}}\relax
\mciteBstWouldAddEndPuncttrue
\mciteSetBstMidEndSepPunct{\mcitedefaultmidpunct}
{\mcitedefaultendpunct}{\mcitedefaultseppunct}\relax
\EndOfBibitem
\bibitem{Abazov:2012bv}
D0, V.~M. Abazov {\em et~al.},
  \ifthenelse{\boolean{articletitles}}{\emph{{Measurement of the W Boson Mass
  with the D0 Detector}},
  }{}\href{http://dx.doi.org/10.1103/PhysRevLett.108.151804}{Phys.\ Rev.\
  Lett.\  \textbf{108} (2012) 151804},
  \href{http://arxiv.org/abs/1203.0293}{{\tt arXiv:1203.0293}}\relax
\mciteBstWouldAddEndPuncttrue
\mciteSetBstMidEndSepPunct{\mcitedefaultmidpunct}
{\mcitedefaultendpunct}{\mcitedefaultseppunct}\relax
\EndOfBibitem
\bibitem{Buge:951371}
V.~Buge {\em et~al.}, \ifthenelse{\boolean{articletitles}}{\emph{{Prospects for
  the Precision Measurement of the W Mass with the CMS Detector at the LHC}},
  }{} Tech. Rep. CMS-NOTE-2006-061, CERN, Geneva, May, 2006\relax
\mciteBstWouldAddEndPuncttrue
\mciteSetBstMidEndSepPunct{\mcitedefaultmidpunct}
{\mcitedefaultendpunct}{\mcitedefaultseppunct}\relax
\EndOfBibitem
\bibitem{Besson:2008zs}
ATLAS, N.~Besson {\em et~al.},
  \ifthenelse{\boolean{articletitles}}{\emph{{Re-evaluation of the LHC
  potential for the measurement of Mw}},
  }{}\href{http://dx.doi.org/10.1140/epjc/s10052-008-0774-4}{Eur.\ Phys.\ J.\
  \textbf{C57} (2008) 627}, \href{http://arxiv.org/abs/0805.2093}{{\tt
  arXiv:0805.2093}}\relax
\mciteBstWouldAddEndPuncttrue
\mciteSetBstMidEndSepPunct{\mcitedefaultmidpunct}
{\mcitedefaultendpunct}{\mcitedefaultseppunct}\relax
\EndOfBibitem
\bibitem{ATL-PHYS-PUB-2014-015}
\ifthenelse{\boolean{articletitles}}{\emph{{Studies of theoretical
  uncertainties on the measurement of the mass of the $W$ boson at the LHC}},
  }{} Tech. Rep. ATL-PHYS-PUB-2014-015, CERN, 2014\relax
\mciteBstWouldAddEndPuncttrue
\mciteSetBstMidEndSepPunct{\mcitedefaultmidpunct}
{\mcitedefaultendpunct}{\mcitedefaultseppunct}\relax
\EndOfBibitem
\bibitem{Krasny:2010vd}
M.~W. Krasny {\em et~al.}, \ifthenelse{\boolean{articletitles}}{\emph{{$\Delta
  M_{W} \leq 10 MeV/c^{2}$ at the LHC: a forlorn hope?}},
  }{}\href{http://dx.doi.org/10.1140/epjc/s10052-010-1417-0}{Eur.\ Phys.\ J.\
  \textbf{C69} (2010) 379}, \href{http://arxiv.org/abs/1004.2597}{{\tt
  arXiv:1004.2597}}\relax
\mciteBstWouldAddEndPuncttrue
\mciteSetBstMidEndSepPunct{\mcitedefaultmidpunct}
{\mcitedefaultendpunct}{\mcitedefaultseppunct}\relax
\EndOfBibitem
\bibitem{Bozzi:2011ww}
G.~Bozzi, J.~Rojo, and A.~Vicini,
  \ifthenelse{\boolean{articletitles}}{\emph{{The Impact of PDF uncertainties
  on the measurement of the W boson mass at the Tevatron and the LHC}},
  }{}\href{http://dx.doi.org/10.1103/PhysRevD.83.113008}{Phys.\ Rev.\
  \textbf{D83} (2011) 113008}, \href{http://arxiv.org/abs/1104.2056}{{\tt
  arXiv:1104.2056}}\relax
\mciteBstWouldAddEndPuncttrue
\mciteSetBstMidEndSepPunct{\mcitedefaultmidpunct}
{\mcitedefaultendpunct}{\mcitedefaultseppunct}\relax
\EndOfBibitem
\bibitem{Rojo:2013nia}
J.~Rojo and A.~Vicini, \ifthenelse{\boolean{articletitles}}{\emph{{PDF
  uncertainties in the extraction of the W mass at LHC: a Snowmass
  Whitepaper}}, }{}\href{http://arxiv.org/abs/1309.1311}{{\tt
  arXiv:1309.1311}}\relax
\mciteBstWouldAddEndPuncttrue
\mciteSetBstMidEndSepPunct{\mcitedefaultmidpunct}
{\mcitedefaultendpunct}{\mcitedefaultseppunct}\relax
\EndOfBibitem
\bibitem{Bozzi:2015hha}
G.~Bozzi, L.~Citelli, and A.~Vicini,
  \ifthenelse{\boolean{articletitles}}{\emph{{PDF uncertainties on the W boson
  mass measurement from the lepton transverse momentum distribution}},
  }{}\href{http://arxiv.org/abs/1501.05587}{{\tt arXiv:1501.05587}}\relax
\mciteBstWouldAddEndPuncttrue
\mciteSetBstMidEndSepPunct{\mcitedefaultmidpunct}
{\mcitedefaultendpunct}{\mcitedefaultseppunct}\relax
\EndOfBibitem
\bibitem{Quackenbush:2015yra}
S.~Quackenbush and Z.~Sullivan,
  \ifthenelse{\boolean{articletitles}}{\emph{{Parton distributions and the $W$
  mass measurement}}, }{}\href{http://arxiv.org/abs/1502.04671}{{\tt
  arXiv:1502.04671}}\relax
\mciteBstWouldAddEndPuncttrue
\mciteSetBstMidEndSepPunct{\mcitedefaultmidpunct}
{\mcitedefaultendpunct}{\mcitedefaultseppunct}\relax
\EndOfBibitem
\bibitem{LHCb-TDR-009}
LHCb collaboration, \ifthenelse{\boolean{articletitles}}{\emph{{LHCb
  reoptimized detector design and performance: Technical Design Report}}, }{}
  \href{http://cdsweb.cern.ch/search?p=CERN-LHCC-2003-030&f=reportnumber&action_search=Search&c=LHCb+Reports}
  {CERN-LHCC-2003-030}.
\newblock LHCB-TDR-009\relax
\mciteBstWouldAddEndPuncttrue
\mciteSetBstMidEndSepPunct{\mcitedefaultmidpunct}
{\mcitedefaultendpunct}{\mcitedefaultseppunct}\relax
\EndOfBibitem
\bibitem{Aaij:1439627}
R.~Aaij {\em et~al.}, \ifthenelse{\boolean{articletitles}}{\emph{{Inclusive $W$
  and $Z$ production in the forward region at $\sqrt{s}$ = 7 TeV}}, }{}J.\ High
  Energy Phys.\  \textbf{06} (2012) 058. 33 p, Comments: 27 pages, 11 figures,
  6 tables\relax
\mciteBstWouldAddEndPuncttrue
\mciteSetBstMidEndSepPunct{\mcitedefaultmidpunct}
{\mcitedefaultendpunct}{\mcitedefaultseppunct}\relax
\EndOfBibitem
\bibitem{Alioli:2008gx}
S.~Alioli, P.~Nason, C.~Oleari, and E.~Re,
  \ifthenelse{\boolean{articletitles}}{\emph{{NLO vector-boson production
  matched with shower in POWHEG}},
  }{}\href{http://dx.doi.org/10.1088/1126-6708/2008/07/060}{JHEP \textbf{0807}
  (2008) 060}, \href{http://arxiv.org/abs/0805.4802}{{\tt
  arXiv:0805.4802}}\relax
\mciteBstWouldAddEndPuncttrue
\mciteSetBstMidEndSepPunct{\mcitedefaultmidpunct}
{\mcitedefaultendpunct}{\mcitedefaultseppunct}\relax
\EndOfBibitem
\bibitem{Sjostrand:2006za}
T.~Sj{\"o}strand, S.~Mrenna, and P.~Z. Skands,
  \ifthenelse{\boolean{articletitles}}{\emph{{PYTHIA 6.4 Physics and Manual}},
  }{}\href{http://dx.doi.org/10.1088/1126-6708/2006/05/026}{JHEP \textbf{0605}
  (2006) 026}, \href{http://arxiv.org/abs/hep-ph/0603175}{{\tt
  arXiv:hep-ph/0603175}}\relax
\mciteBstWouldAddEndPuncttrue
\mciteSetBstMidEndSepPunct{\mcitedefaultmidpunct}
{\mcitedefaultendpunct}{\mcitedefaultseppunct}\relax
\EndOfBibitem
\bibitem{Ball:2014uwa}
NNPDF, R.~D. Ball {\em et~al.},
  \ifthenelse{\boolean{articletitles}}{\emph{{Parton distributions for the LHC
  Run II}}, }{}\href{http://dx.doi.org/10.1007/JHEP04(2015)040}{JHEP
  \textbf{1504} (2015) 040}, \href{http://arxiv.org/abs/1410.8849}{{\tt
  arXiv:1410.8849}}\relax
\mciteBstWouldAddEndPuncttrue
\mciteSetBstMidEndSepPunct{\mcitedefaultmidpunct}
{\mcitedefaultendpunct}{\mcitedefaultseppunct}\relax
\EndOfBibitem
\bibitem{Harland-Lang:2014zoa}
L.~A. Harland-Lang, A.~D. Martin, P.~Motylinski, and R.~S. Thorne,
  \ifthenelse{\boolean{articletitles}}{\emph{{Parton distributions in the LHC
  era: MMHT 2014 PDFs}},
  }{}\href{http://dx.doi.org/10.1140/epjc/s10052-015-3397-6}{Eur.\ Phys.\ J.\
  \textbf{C75} (2015), no.~5 204}, \href{http://arxiv.org/abs/1412.3989}{{\tt
  arXiv:1412.3989}}\relax
\mciteBstWouldAddEndPuncttrue
\mciteSetBstMidEndSepPunct{\mcitedefaultmidpunct}
{\mcitedefaultendpunct}{\mcitedefaultseppunct}\relax
\EndOfBibitem
\bibitem{CT10}
J.~Gao {\em et~al.}, \ifthenelse{\boolean{articletitles}}{\emph{{CT10
  next-to-next-to-leading order global analysis of QCD}}, }{}Phys.\ Rev.\ D
  \textbf{89} (2014) 033009\relax
\mciteBstWouldAddEndPuncttrue
\mciteSetBstMidEndSepPunct{\mcitedefaultmidpunct}
{\mcitedefaultendpunct}{\mcitedefaultseppunct}\relax
\EndOfBibitem
\bibitem{Botje:2011sn}
M.~Botje {\em et~al.}, \ifthenelse{\boolean{articletitles}}{\emph{{The PDF4LHC
  Working Group Interim Recommendations}},
  }{}\href{http://arxiv.org/abs/1101.0538}{{\tt arXiv:1101.0538}}\relax
\mciteBstWouldAddEndPuncttrue
\mciteSetBstMidEndSepPunct{\mcitedefaultmidpunct}
{\mcitedefaultendpunct}{\mcitedefaultseppunct}\relax
\EndOfBibitem
\bibitem{Aaij:2015gna}
R.~Aaij {\em et~al.}, \ifthenelse{\boolean{articletitles}}{\emph{{Measurement
  of the forward $Z$ boson production cross-section in $pp$ collisions at
  $\sqrt{s}$ = 7 TeV}}, }{}\href{http://arxiv.org/abs/1505.07024}{{\tt
  arXiv:1505.07024}}\relax
\mciteBstWouldAddEndPuncttrue
\mciteSetBstMidEndSepPunct{\mcitedefaultmidpunct}
{\mcitedefaultendpunct}{\mcitedefaultseppunct}\relax
\EndOfBibitem
\bibitem{Aaij:2013qja}
R.~Aaij {\em et~al.}, \ifthenelse{\boolean{articletitles}}{\emph{{Measurement
  of the $\Lambda_b^0$, $\Xi_b^-$ and $\Omega_b^-$ baryon masses}},
  }{}\href{http://dx.doi.org/10.1103/PhysRevLett.110.182001}{Phys.\ Rev.\
  Lett.\  \textbf{110} (2013), no.~18 182001},
  \href{http://arxiv.org/abs/1302.1072}{{\tt arXiv:1302.1072}}\relax
\mciteBstWouldAddEndPuncttrue
\mciteSetBstMidEndSepPunct{\mcitedefaultmidpunct}
{\mcitedefaultendpunct}{\mcitedefaultseppunct}\relax
\EndOfBibitem
\bibitem{Aaij:2013uaa}
R.~Aaij {\em et~al.}, \ifthenelse{\boolean{articletitles}}{\emph{{Precision
  measurement of D meson mass differences}},
  }{}\href{http://dx.doi.org/10.1007/JHEP06(2013)065}{JHEP \textbf{1306} (2013)
  065}, \href{http://arxiv.org/abs/1304.6865}{{\tt arXiv:1304.6865}}\relax
\mciteBstWouldAddEndPuncttrue
\mciteSetBstMidEndSepPunct{\mcitedefaultmidpunct}
{\mcitedefaultendpunct}{\mcitedefaultseppunct}\relax
\EndOfBibitem
\bibitem{Aaij:1428520}
R.~Aaij {\em et~al.}, \ifthenelse{\boolean{articletitles}}{\emph{{Measurement
  of $\Upsilon$ production in $pp$ collisions at $\sqrt{s} = 7$ TeV}}, }{}Eur.\
  Phys.\ J.\ C \textbf{72} (2012) 2025. 21 p\relax
\mciteBstWouldAddEndPuncttrue
\mciteSetBstMidEndSepPunct{\mcitedefaultmidpunct}
{\mcitedefaultendpunct}{\mcitedefaultseppunct}\relax
\EndOfBibitem
\bibitem{Landry:2002ix}
F.~Landry, R.~Brock, P.~M. Nadolsky, and C.~P. Yuan,
  \ifthenelse{\boolean{articletitles}}{\emph{{Tevatron Run-1 $Z$ boson data and
  Collins-Soper-Sterman resummation formalism}},
  }{}\href{http://dx.doi.org/10.1103/PhysRevD.67.073016}{Phys.\ Rev.\
  \textbf{D67} (2003) 073016}, \href{http://arxiv.org/abs/hep-ph/0212159}{{\tt
  arXiv:hep-ph/0212159}}\relax
\mciteBstWouldAddEndPuncttrue
\mciteSetBstMidEndSepPunct{\mcitedefaultmidpunct}
{\mcitedefaultendpunct}{\mcitedefaultseppunct}\relax
\EndOfBibitem
\bibitem{Konychev:2005iy}
A.~V. Konychev and P.~M. Nadolsky,
  \ifthenelse{\boolean{articletitles}}{\emph{{Universality of the
  Collins-Soper-Sterman nonperturbative function in gauge boson production}},
  }{}\href{http://dx.doi.org/10.1016/j.physletb.2005.12.063}{Phys.\ Lett.\
  \textbf{B633} (2006) 710}, \href{http://arxiv.org/abs/hep-ph/0506225}{{\tt
  arXiv:hep-ph/0506225}}\relax
\mciteBstWouldAddEndPuncttrue
\mciteSetBstMidEndSepPunct{\mcitedefaultmidpunct}
{\mcitedefaultendpunct}{\mcitedefaultseppunct}\relax
\EndOfBibitem
\bibitem{Berge:2005rv}
S.~Berge, P.~M. Nadolsky, and F.~I. Olness,
  \ifthenelse{\boolean{articletitles}}{\emph{{Heavy-flavor effects in soft
  gluon resummation for electroweak boson production at hadron colliders}},
  }{}\href{http://dx.doi.org/10.1103/PhysRevD.73.013002}{Phys.\ Rev.\
  \textbf{D73} (2006) 013002}, \href{http://arxiv.org/abs/hep-ph/0509023}{{\tt
  arXiv:hep-ph/0509023}}\relax
\mciteBstWouldAddEndPuncttrue
\mciteSetBstMidEndSepPunct{\mcitedefaultmidpunct}
{\mcitedefaultendpunct}{\mcitedefaultseppunct}\relax
\EndOfBibitem
\end{mcitethebibliography}

\end{document}